\documentclass{aastex63}
\usepackage{color}
\usepackage{slantsc}

\newcommand\new[1]{\textcolor{red}{{#1}}}

\received{January 7, 2021}
\submitjournal{ApJS}

\shorttitle{Oxygen Cation Spectra}
\shortauthors{Fortenberry, Bodewits, \& Pierce}

\begin{document}

\title{Knowledge Gaps in the Cometary Spectra of Oxygen-Bearing Molecular Cations}

\correspondingauthor{Ryan C. Fortenberry}
\email{r410@olemiss.edu}

\author[0000-0003-4716-8225]{Ryan C. Fortenberry}
\affiliation{Department of Chemistry \& Biochemistry, University of Mississippi, University, MS 38677-1848, USA}

\author[0000-0002-2668-7248]{Dennis Bodewits}
\affiliation{Department of Physics,  Auburn University, Leach Science Center, Auburn, AL 36832, USA}

\author[0000-0002-3946-780X]{Donna M. Pierce}
\affiliation{Department of Physics and Astronomy, Mississippi State University, P.O. Box 5167, Mississippi State, MS 39762-5167, USA}



\begin{abstract}

Molecular cations are present in various astronomical environments, most notably in cometary atmospheres and tails where sunlight produces exceptionally bright near-UV to visible transitions. Such cations typically have longer-wavelength and brighter electronic emission than their corresponding neutrals. A robust understanding of their near-UV to visible properties would allow these cations to be used as tools for probing the local plasma environments or as tracers of neutral gas in cometary environments. However, full spectral models are not possible for characterization of small, oxygen containing molecular cations given the body of molecular data currently available.   The five simplest such species (H$_2$O$^+$, CO$_2^+$, CO$^+$, OH$^+$, and O$_2^+$) are well characterized in some spectral regions but are lacking robust reference data in others. Such knowledge gaps hinder fully quantitative models of cometary spectra, specifically, hindering accurate estimates of physical-chemical processes originating with the most common molecules in comets.  Herein the existing spectral data are collected for these molecules and highlight the places where future work is needed, specifically where the lack of such data would greatly enhance the understanding of cometary evolution. 

\end{abstract}

\keywords{molecular spectroscopy --- catalogs --- surveys}


\section{Introduction} \label{sec:intro}

Ongoing improvements in detection limits and spectroscopic resolution in various wavelength regions have led to the discovery of many molecular and atomic emission lines in myriad astrophysical environments \citep{McGuire2018}, not the least of which are comets \citep{feldman2005,bm2017}. However, there are still thousands of unidentified cometary lines from the near-UV to visible with many likely resulting from unknown transitions of known molecules that cloud spectral classification \citep{brown96, morrison97, wyckoff1999, mumma2001, cochran02, cremonese2007, kawakita02, dr2013, opitom19}. Small molecular ions, such as H$_2$O$^+$, CO$_2^+$, CO$^+$, OH$^+$, and O$_2^+$, have been observed in a variety of various astronomical environments, including comet comae and tails, planetary atmospheres, planetary disks, and the interstellar medium (ISM) \citep{Larsson2012}.  They are often sentinel species for underlying photochemical processes which are driven in our Solar System by the Sun. Molecular radical cations also typically produce emission features at longer wavelengths than those observed from closed-shell, neutral molecules (often in the visible, whereas neutrals are in the UV). Accurate and complete spectra, if publicly available, could be used to determine if currently unidentified lines observed in cometary spectra are due to molecular cations, as exemplified by the successful attribution of previously unidentified spectral lines in comets Hyakutake and Ikeya-Zhang \citep{wyckoff1999, kawakita02} to transitions from higher-excited levels of H$_2$O$^+$ \citep{Bodewits2016}.  Such a database would also provide a means of quantifying the molecular cations such that ionization rates, sublimation and mass loss rates, local plasma conditions, and potentially even cometary chemical classifications would be enhanced \citep{Raghuram2021}.  To derive column densities and chemical abundances, heliocentric velocity-dependent fluorescence efficiencies of H$_2$O$^+$, CO$_2^+$, CO$^+$, and OH$^+$ are needed.  While such values are known in part for some of these molecules, the full set of fluorescence data are lacking in many, most notably in H$_2$O$^+$, and these knowledge gaps should be filled. 

The remote identification of  ions, especially in comets, requires knowledge of their spectra. The analysis and interpretation of ion data requires spectral models based on well-characterized values for dipole moments, line positions, Einstein A and B coefficients, and fluorescence efficiency factors. Surprisingly, this information is not readily available. Spectral data for these molecules are mostly missing from major databases, such as the Cologne Database for Molecular Spectroscopy  \citep[CDMS; ][]{muller2001}, the high-resolution transmission molecular absorption database \citep[HITRAN; ][]{Gordon2017}, the Virtual Planetary Laboratory (VPL) Molecular Spectroscopic Database\footnote{\url{http://vpl.astro.washington.edu/spectra/fundamentals.htm}}, and the Diatomic and Triatomic spectral databases at NIST \citep{NIST2,NIST3}. Most of these databases are predominantly geared towards applications in the IR and microwave regions, and near-UV to visible rovibronic spectra are not as well characterized.  As a consequence, many currently unidentified lines can likely be attributed to these and other ions, as is the case in comet spectra \citep{wyckoff1999, kawakita02, bodewits2019}, but these lines go unattributed due to a lack of the mentioned reference data.

The current hindrance of comet studies due to the lack of spectral data will be illustrated with three examples.

 First, the near-UV to visible spectrum of the unusual comet C/2016~R2 (PanSTARRS) was mostly dominated by CO$^+$ emission, with barely any evidence of water or its fragments \citep{mckay19, opitom19, venkat20}. Comparison with detailed spectral models would allow observers to carefully remove the emission of CO$^+$, which obscures the possible emission of other species \citep{venkat20}.  \citet{cochran18} obtained robust spectra of CO$^{+}$ in the tail of comet C/2016~R2~(Pan-STARRS) when it was more than 3 AU from the sun, and these observations also yielded a rare, unequivocal cometary detection of N$_{2}^{+}$. Prior to this result, most spectroscopic detections of N$_{2}^{+}$ in comet spectra are thought to be telluric lines.  The N$_{2}^{+}$/CO$^+$ ratio often serves as a substitute for the N$_{2}$/CO ratio in comets. The N$_2$/CO abundance ratio may reflect how cometary ices formed \citep{Rubin2015}.  Other species may also be masked under similar circumstances and more data are needed to feed the models for comparison.
 
Second, one of the most surprising results of the {\em Rosetta} mission to comet 67P/Churyumov-Gerasimenko was the detection of high amounts of O$_2$ in its coma (up to 10\%), which has a pronounced impact on planet formation models \citep{Bieler15, keeney17}. Molecular oxygen has no dipole moment and thus produces no emission from its rotational transitions. Its near-UV to visible emission features are dipole forbidden and likely weak if present \citep{Glinski2004}. On the other hand, its cation, which has been studied only at low resolution for its electronic emission as discussed later, has several emission lines in the visible regime (Table 1) that are currently lacking in full calibration. Reliable spectral models would allow observers to search for the remote emission of O$_2^+$ in comets and other small bodies.

Third, CO$_2$ is one of the main volatiles in comets \citep{bm2017}, and its abundance relative to H$_2$O and CO may link comets to their formation regions \citep{ahearn2012}. Unfortunately, the emission of CO$_2$ is blocked by the Earth’s atmosphere. However, CO$_2^+$ produces bright lines in the near-UV ($B \rightarrow X$), and has been observed in many comets \citep[cf.][]{weaver81}, as well as the less frequently observed $ A \rightarrow X $ transition at longer wavelengths \citep{opitom19}. Observers interested in characterizing the activity and volatile content of comets would benefit greatly from digitized, electronically-available high-resolution spectra along with accurate fluorescence rates that are currently not available largely due to a lack of high-resolution experimental or theoretical results.

This paper presents a critical analysis for the state of spectroscopic data currently available (Table \ref{overview}), as well as existing data gaps, for diatomic and triatomic oxygen-bearing cations that are abundant in comets:~~H$_2$O$^+$, CO$_2^+$, CO$^+$, OH$^+$, and O$_2^+$.  These five cations are produced by photodissociation or electron-impact of the most common cometary neutral species:~~H$_2$O, CO$_2$, and CO \citep{bm2017, Beth2020}. Our discussion is limited to transitions above 200 nm, roughly the lower limit for the quantum efficiency of most CCD detectors requiring different hardware for observation. These knowledge gaps exist due to various factors ranging from difficulty of the experiments to simply a lack of motivation to study these systems.  In any case, the following section describes what is currently known and offers a call to determine the remaining factors in order to produce much more insightful spectral models of comets or any other astrophysical objects where these molecules may be found.

\section{Review of existing spectroscopic models}

\begin{table}
\begin{center}
\caption{\label{overview} Summary of the main emission features of small molecular cations applicable to rovibronic modeling of comet spectra with recommended references.}   
\begin{tabular}{lclcc}
{\bf Species}   & {\bf Transition/System} & {\bf Name} & {\bf Wavelength}  & {\bf References}    \\
\hline                    
H$_2$O$^+$ & $A\ ^2A_1 \rightarrow X\ ^2B_1$ &  & 4200 -- 7500 \AA & \citet{Lew73, Lew76} \\
H$_2$O$^+$ & $B\ ^2B_2 \rightarrow X\ ^2B_1$ &  & 2200 -- 2800 \AA & \citet{Reutt86} \\
H$_2$O$^+$ & $B\ ^2B_2 \rightarrow A\ ^2A_1$ &  & 3000 -- 4300 \AA & \citet{Reutt86} \\

CO$_2^+$ & $A\ ^2\Pi_u \rightarrow X\ ^2\Pi_g$ & Fox-Duffendack-Barker & 2800 - 5000 \AA &    \citet{Fox27, Gauyacq75} \\
CO$_2^+$ & $B\ ^2\Sigma_u \rightarrow X\ ^2\Pi_g$&& 2890 \AA & \citet{Gauyacq75}     \\
CO$_2^+$ & $C\ ^2\Sigma_g \rightarrow X\ ^2\Pi_g$&& 2215 \AA & \citet{Wang88, Wyttenbach89}     \\

CO$^+$   & $B\ ^2\Sigma^+ \rightarrow X\ ^2\Sigma^+$ &  First Negative &2100 -- 2500 \AA  & \citet{judge72, lee74}   \\
CO$^+$   & $B\ ^2\Sigma \rightarrow A\ ^2\Pi$ &  Baldet-Johnson & 3620 -- 6165 \AA  & \citet{dotchin73}   \\
CO$^+$   & $A\ ^2\Pi \rightarrow X\ ^2\Sigma^+$ & Comet tail &  4000 -- 6000 \AA  & \citet{judge72}    \\

OH$^+$   & $A\ ^3\Pi \rightarrow X\ ^3\Sigma^-$  & & 3300 -- 3600 \AA & \citet{Hodges2017}       \\

O$_2^+$  &  $A\ ^2\Pi_u \rightarrow X\ ^2\Pi_g$  &Second Negative&  1800 -- 5300 \AA   & \citet{Terrell2004}  \\
O$_2^+$  &  $b\ ^4\Sigma_g^- \rightarrow a\ ^4\Pi_u$  &First Negative & 4500 -- 8500 \AA & \citet{Terrell2004,Glinski2004}   \\
\hline
\multicolumn{5}{l}{{\bf Note:} Where applicable, we follow the bent notation for the vibronic structure \citep[cf.][]{cochran02}.}
\end{tabular}

\end{center}
\end{table}

\subsection{H$_2$O$^+$}

The water cation first made its presence felt in planetary science in 1974 when the $A\ ^2A_1 \rightarrow X\ ^2B_1$ emission was tentatively observed nearly simultaneously by two groups \citep{Benvenuti74, Herzberg74} at near-UV to visible wavelengths in the tail of comet Kohoutek (1973f or C/1973 E1).  Additionally, \citet{wehinger74b} made a detection of the same transition in the tail of comet Bradfield (1974b = C/1974 C1) shortly thereafter, and follow-up work confirmed the presence of H$_2$O$^+$ in Kohoutek \citep{Wehinger74}, opening the door for observations in other comets.  Thirteen other comets showed spectral signatures of H$_2$O$^+$ within the next few years \citep{Miller80}. The high-resolution spectral atlas by \citet{cochran02} contains an inventory of 129 observed lines of 3 vibrational transitions of H$_2$O$^+$ spanning the wavelength range of 5800 -- 7500 \AA\ in comet 122P/De Vico. \cite{bodewits2019} used electron impact-induced spectra to attribute numerous unidentified lines in the ion tail of comet Hyakutake, reported by \citet{wyckoff1999} and \citet{kawakita02}, to transitions from higher vibrational levels of the H$_2$O$^+$  $A\ ^2A_1 \rightarrow X\ ^2B_1$ transition.  Additionally, \citet{lutz87} calculated fluorescence efficiency factors and absorption oscillator strengths for the same transition of H$_{2}$O$^{+}$ in order to determine its abundance in the tail of comet Kohoutek.

Even with this relatively early detection of H$_2$O$^+$ in comets, it was not until 2010 before H$_2$O$^+$ was observed in the ISM via radiotelescopic detection towards various star-forming regions \citep{Ossenkopf10}, and H$_{2}$O$^{+}$ has now even been observed in older, extragalactic objects with $Z>0.8$ with ALMA \citep{Muller16}.  In fact, this work exploring spectral features at such high redshift found that the existing rotational spectral data for $X\ ^2B_1$ H$_2$O$^+$ at the time were not of sufficient quality for more detailed modeling to be performed \citep{Muller16}.  The lines in the CDMS \citep{muller2001} matched the observations adequately in order to claim a detection, but the new rest frequencies reported in CDMS are actually those from ALMA observations combined with laboratory work done as a part of the \cite{Muller16} study.  These are listed in Table \ref{h2o+} for the $X\ ^2B_1$ $\nu=0$ state. Consequently, the pure rotational frequencies and spectroscopic constants of H$_2$O$^+$ are established to high precision.

The three fundamental vibrational frequencies for $X\ ^2B_1$ of H$_2$O$^+$  in the gas phase have been determined through photoelectron and laser difference spectroscopy at 3212.86 cm$^{-1}$, 1408.42 cm$^{-1}$, and 3259.04 cm$^{-1}$ \citep{Reutt86, Jacox94}, as shown in Table \ref{h2o+}.  Refinements to these values were made by \cite{Muller16}, extending the number of significant figures.  Additionally, this refinement also generated rovibrational spectroscopic constants for these singly-excited/one quantum fundamental vibrational frequencies \citep{Muller16} allowing for accurate rovibrational models of H$_2$O$^+$ to be produced.  However, the electronic and vibronic bands are still incomplete.

The electronic spectral characterization of H$_2$O$^+$ coincided with its initial observation in comet Kohoutek.  Work by \cite{Lew73} and \cite{Lew76} provides clear descriptions of the electronic feature for the $A\ ^2A_1 \rightarrow X\ ^2B_1$ transition, but the full vibronic nature of any excited states have yet to be established.  The $A\ ^2A_1$ and $B\ ^2B_2$ states have accurate transition energies from photoelectron experiments \citep{Reutt86}, putting the $B\ ^2B_2 \rightarrow X\ ^2B_1$ transition in the 2200 -- 2800 \AA\ range, and the $B\ ^2B_2 \rightarrow A\ ^2A_1$ transition in the 3000 -- 4300 \AA~range, as given in Table \ref{overview}.  These ranges, while largely estimated for the two transitions involving the $B\ ^2B_2$ state, are well-separated, giving H$_2$O$^+$ full emission coverage from 2200 \AA\ through the UV to the very edge of the visible regime at 7500 \AA.  More resolved electronic transition wavelengths to at least the sub nm scale, and certainly more accurate and complete vibrational levels to at least the single cm$^{-1}$ scale for these electronically excited states of H$_2$O$^+$, are needed in order to model these spectra more effectively.  Additionally, more narrowly defined state lifetimes (and, hence, Einstein coefficients) are also needed as the current measurements are more than 40 years old and two different studies disagree with one another by nearly a factor of three \citep{Curtis77, Mohlmann78}.

\begin{table}
\begin{center}
\caption{\label{h2o+} Reported Vibrational Frequencies (cm$^{-1}$) and Rotational Constants (MHz) for the Various Electronic States of H$_2$O$^+$}   
\begin{tabular}{llcccc}
State  & & $\nu=0$ & $\nu_1=1$ ($a_1$) & $\nu_2=1$ ($a_1$) & $\nu_3=1$ ($b_2$)   \\
\hline                    
$X\ ^2B_1$$^a$ & $h\nu$ & 0.0 & 1408.4131 & 3212.8567 & 3259.0341 \\
 & $A$       & 870 580.8 & 1 001 285. & 851 254. & 835 041. \\
 & $B$       & 372 365.4 & 374 077.4 & 365 511.7 & 367 803.7 \\
 & $C$       & 253 880.4 & 249 275.7 & 248 680.5 & 349 733.7 \\
 & $\Delta_K$ & 1375.3 & 2902.6 & 1348.5 & 1269.2 \\
 & $\Delta_{NK}$& -155.30 & -246.4 & -154.6 & -158.6 \\
 & $\Delta_N$ & 29.66 & 31.00 & 29.80 & 30.23 \\
\hline
$A\ ^2A_1$$^b$ & $h\nu$ & 13409.3 & 3547 & 876.8 & \\
$B\ ^2B_2$$^b$ & $h\nu$ & 36757   & 2968 & 1596  & \\

\hline
\end{tabular}
\end{center}
$^a$All rovibrational data for the $X\ ^2B_1$ state are from \cite{Muller16} which further includes $\delta$, $\Phi$, $\phi$, $L$, and $\epsilon$ values.\\
$^b$The vibronic fundamentals are compiled from \cite{Lew76, Reutt86, Jacox94}.  See text for discussion.\\
\end{table}

The experimental photoelectron results given in Table \ref{h2o+} only produce estimates for fundamental vibrational frequencies, and only for the $a_1$ modes \citep{Reutt86}.  The $\nu_2$ bend of the $A\ ^2A_1$ state at 876.8 cm$^{-1}$ is the most concretely established fundamental frequency based on comparisons between the photoelectron data and the prior existing emission spectra \citep{Lew76, Reutt86, Jacox94}.  These values guided electron impact studies that established many of the experimental rovibronic bands for the $\nu_2$ bend of the $A\ ^2A_1$ state \citep{Kuchenev96}.  Subsequent quantum chemical vibronic lines in the $A\ ^2A_1 \rightarrow X\ ^2B_1$ transition were compared to those from the electron impact study.  The error is less than 2 cm$^{-1}$ for this purely theoretical study \citep{Wu04}, providing high-accuracy vibronic lines for higher quanta of the $A\ ^2A_1$ $\nu_2$ bend.  These, in turn, were subsequently corroborated and extended experimentally \citep{Gan04}.  Even so, no lines or spectroscopic constants for higher quanta in either of the other two modes of $A\ ^2A_1$, or for any modes of the $B\ ^2B_2$ state, are currently given in the literature.  Noble gas tagging experiments have produced estimates for the hydride stretches, but the proton-bound nature of these complexes likely makes them irrelevant for reference data required of H$_2$O$^+$ \citep{Roth01, Dopfer01}.  

Consequently, the high-accuracy spectral data (both experimental and theoretical) available for the water cation from the literature currently consist of:
\begin{itemize}
    \item Pure rotational spectroscopic constants and lines available in CDMS,
    \item Fundamental vibrational frequencies and rovibrational spectroscopic constants that can be produced from \cite{Muller16}, 
    \item Onset energies for the $A\ ^2A_1 \rightarrow X\ ^2B_1$, $B\ ^2B_2 \rightarrow X\ ^2B_1$, and $B\ ^2B_2 \rightarrow A\ ^2A_1$ transitions \citep{Lew73, Lew76, Reutt86, Jacox94}
    \item The $\nu_2$ bands of the $A\ ^2A_1$ state provided by \cite{Wu04}.
\end{itemize}
The remaining data necessary for proper cometary or other astrochemical modeling are the rovibrational spectroscopic constants and lines for the $A\ ^2A_1$ and $B\ ^2B_2$ excited states, save for the $\nu_2$ bands of the $A\ ^2A_1$ state as established by \citet{Kuchenev96}, \citet{Wu04}, and \citet{Gan04}.  These electronic states lie at 7457.7 \AA\ and 2708.7 \AA\ above the ground state \citep{Lew73, Lew76, Reutt86, Jacox94}, making them the only two states of interest for purely solar excitation models like those for comets.  Hence, full modeling of H$_2$O$^+$ in comets will require line lists for these two electronic states, which are currently unavailable from any previous experimental or theoretical studies.  Finally, Einstein coefficients for transitions involving each of these states are also needed in order for the spectral models to be complete.

\subsection{CO$_2^+$}

The detection of the carbon dioxide cation predates that of H$_2$O$^+$ by nearly 25 years. Observations of the long-period comet Bester (1947k or C/1947 F1) acquired as it was approaching the sun revealed numerous bands of the CO$_2^+ \widetilde{A}\ ^2\Pi_u \rightarrow \widetilde{X}\ ^2\Pi_g$ emission transition in the comet tail at wavelengths from 3500 \AA\ to 5000 \AA\ \citep{Swings50}. The spectrum of this molecule had been known since the late 1920s \citep{Fox27, Duffendack29} and was refined in the early 1940s \citep{Mrozowski41}, but comet Bester provided the first natural observation of CO$_2^+$ beyond the Earth's atmosphere.  Both CO$_2^+$ $\widetilde{A}\ ^2\Pi_u \rightarrow \widetilde{X}\ ^2\Pi_g$ and the $\widetilde{B}\ ^2\Sigma^+_u \rightarrow \widetilde{X}\ ^2\Pi_g$ transitions were observed by {\em Mariner 6} and {\em Mariner 7}~in the dayglow of Mars \citep{Barth71}.  Sounding rocket observations of comet C/1975 V1 (West) yielded the first cometary detection of the CO$_2^+ \widetilde{B}\ ^2\Sigma^+_u \rightarrow \widetilde{X}\ ^2\Pi_g$ transition in the near-UV \citep{Feldman1976}. The detection of CO$_2^+$ also predates the first cometary observation of the neutral by 36 years, when its fluorescence spectrum was observed by the \emph{Vega 1} probe to comet 1P/Halley in 1986 \citep{Combes86}. Later, the emission of neutral CO$_2$ was first discovered in the ISM via IR spectra in 1989 via the bright antisymmetric stretch mode \citep{Hendecourt89}, but the cation has yet to be documented in any environment beyond the Solar System.

\begin{table}
\begin{center}
\caption{\label{co2+} Reported Vibrational Frequencies (cm$^{-1}$) and Rotational Constants (MHz) for the Various Electronic States of CO$_2^+$.}   
\begin{tabular}{llcccc}
State  & & $\nu=0$ & $\nu_1=1$ ($\sigma_g^+$) & $\nu_2=1$ ($\pi_g$) & $\nu_3=1$ ($\sigma_u^+$)   \\
\hline                    
$X\ ^2\Pi_g$$^a$ & $h\nu$ & 0.0 & 1264.98 & 513.262 & 1423.08 \\
 & $B$       & 11 410 &  & 11 380  &  \\
 & $D$       & 0.033 &  &  &  \\
\hline
$A\ ^2\Pi_u$$^a$ & $h\nu$ & 28500.35 & 1122.41 & 466.75 & 2653.07\\
 & $B$       & 10 500 &  &  &  \\
 \hline
$B\ ^2\Sigma_u^+$$^b$& $h\nu$ & 34591.6  & 1284 & 590 & 1891 \\
 & $B$       & 11 340 &  & 11 500 &  \\
 & $D$       & 0.069 &  &  &  \\
 \hline
$C\ ^2\Sigma_g^+$$^c$& $h\nu$ & 45157  & 1384 & 614 & 1567 \\
 & $B$       & 11 837 & 11 820 & &  \\
 & $D$       & 0.1 & 0.1 &  &  \\  
  
\hline
\end{tabular}

\end{center}
$^a$ Data collected from \cite{Gauyacq75, Varfalzy07, Gharaibeh10, Jacox03}.\\
$^b$ Data collected from \cite{Gauyacq75, Jacox03}.\\
\new{$^c$} Data collected from \cite{Wang88, Wyttenbach89, Jacox03}.
\\

\end{table}

The vibronic spectrum of CO$_2^+$ is very well established, even if very complicated \citep{Gauyacq75}.  This molecular spectrum is known primarily due to its presence in comet Halley \citep{Johnson95}. As referenced in Table \ref{co2+}, high precision experimental results are available for the ground state ($X\ ^2\Pi_g$) and first excited state ($A\ ^2\Pi_u$ at 3508.73 \AA), along with detailed, though less precise, data for the higher $B\ ^2\Sigma_u^+$ (2890.88 \AA) and $C\ ^2\Sigma_g^+$ (2214.88 \AA) states \citep{Jacox03, McCallum72, Frye87, Wang88, Wyttenbach89, Johnson95, Chambaud92, Liu00, Shaw05, Gharaibeh10, NIST3}. These higher states, especially the $C\ ^2\Sigma_g^+$, also lie beyond what can be excited through photofluorescence, the dominant emission mechanism for most cometary fragment species \citep{Shaw05}.  \cite{Johnson95} provide a thorough review of the state of CO$_2^+$ spectral data that were available at the time of their publication.  This molecule has problematic Fermi resonances, Renner-Teller distortions, and vibronic couplings \citep{Varfalzy07}; all of which shift the spectral lines from the locations predicted by the standard model and put vibrational transitions of electronically excited states in competition with other features, making them difficult to be clearly delineated. 

However, \cite{Chambaud92} report an extensive,  experimentally-benchmarked list of theoretically-computed vibronic transitions within the 5000 cm$^{-1}$ of the $X\ ^2\Pi_g$ ground state for CO$_2^+$, where the complicated interactions are mitigated through control of the quantum chemical computations employed.  These values have been corroborated and extended more recently by \cite{Liu00} and \cite{Gharaibeh10}, among others, through pulsed field ionization–photoelectron and laser-induced fluorescence, respectively, to more than 10000 cm$^{-1}$ beyond the onset of the $A\ ^2\Pi_u$ state.  Rotational and spectroscopic constants have also been provided for the fundamental frequencies of the IR-allowed $\nu_3$ mode of the ground electronic state, as well as the $\nu_1$ and $\nu_3$ modes of the $A\ ^2\Pi_u$ state via high-precision emission spectroscopy \citep{Johnson95}.  Similar to its neutral counterpart, CO$_2^+$ does possess the bright $\nu_3$ antisymmetric stretching mode, but the intensity of this mode for the cation is reduced by approximately 75\% compared to its neutral counterpart according to double-harmonic quantum chemical computations performed presently.

Photofluorescence efficiencies for the most important bands in the $A\ ^2\Pi_u \rightarrow X\ ^2\Pi_g$ and $B\ ^2\Sigma_u^+ \rightarrow X\ ^2\Pi_g$ transitions have been determined by \cite{Kim99}, using the transition probability data of \citet{mccallum1971}, and the results were compared to ground-based spectra of comets Austin (1989c1 = C/1989 X1) and 21P/Giacobini-Zinner. \cite{Kim99} argued that the Swings effect did not significantly affect the fluorescence efficiency of the CO$_2^+$ transitions and that the accuracy of their model was mostly limited by spectroscopic data  (line positions and transition probabilities) available to them.  Furthermore, their equilibrium model only includes vibrational-vibrational transitions, unlike the aforementioned studies in the prior paragraph and others \citep{Itikawa2001}, which cover rovibrational transitions and focus heavily on the Renner-Teller effect. However, as demonstrated in the discussion above, significant work has been done to understand the spectrum of CO$_{2}^{+}$ after their study was published, thus meriting a re-evaluation of the data.  Consequently, a majority of the necessary spectral data to model CO$_2^+$ should already be available from these references, which are largely curated at NIST \citep{NIST3}.  The exceptions are the vibrationally-excited rotational constants of the excited electronic states and absolute line intensities (oscillator strengths) for these transitions, as most currently available are relative at best.  While not as necessary for the production of the electronic band progressions, having the full rovibronic scope of data, Einstein coefficients, and the absolute intensities would produce the most accurate and descriptive models for comparison to observation.

The $C\ ^2\Sigma_g^+$ state has not been observed in comets even though excitations into the longer wavelength $B\ ^2\Sigma_u^+$ state have been observed from space-based instruments \citep{weaver81}.  However, some rovibrational data for this state are known with fairly high resolution \citep{Wyttenbach89}, especially the excitation energy and the rotational constants.  However, the rotational quartic distortion $D$ constant was held fixed in the work by \cite{Wyttenbach89} for each vibrational state.  \cite{Wang88} and \cite{Wyttenbach89} disagree as to the assignment of the $\nu_1$ symmetric stretching frequency (1352 cm$^{-1}$ versus 1384 cm$^{-1}$, respectively), but the latter also reports the $B$ rotational constant for this mode lending more credence to their assignment.  The other fundamental vibrational frequencies are from \cite{Wang88}, and the rotational data for these modes are also missing.  Hence, the $C\ ^2\Sigma_g^+$ state of CO$_2^+$ could also benefit from more advanced study.  Such data may be able to isolate this state in cometary spectra which is in a region that is difficult to observe.  Such observations, however, can only occur if more data are on hand for fluorescence models to compare with observation.

Transitions between the various excited electronic states of CO$_2^+$ are also possible originating from solar radiation.  However, most are at near-infrared wavelengths and would likely have relatively low molecular state populations involved in such transitions giving little signal.  Additionally, such weak signals also create challenges for producing high-resolution experimental spectral data in the laboratory.  Regardless, the $B\ ^2\Sigma_u^+ \rightarrow A\ ^2\Pi_u$ transition should take place at 16417 \AA\ based on the difference in the electronic energies for each of these states from \cite{Gharaibeh10} and \cite{Johnson95}.  The $C\ ^2\Sigma_g^+ \rightarrow A\ ^2\Pi_u$ transition wavelength is actually in the visible region at 6003.6 \AA\ \citep{Wyttenbach89} leaving the $C\ ^2\Sigma_g^+ \rightarrow B\ ^2\Sigma_u^+$ in between these two at 9464.9 \AA.  The photofluorescence models of CO$_2^+$ would also benefit from data for these excited states even if they would be minor contributors to the overall cometary spectra of interest.

\subsection{CO$^+$}

The detection of the  $A\ ^{2} \Pi_{i} \rightarrow X\ ^{2} \Sigma^{+}$ transition of CO$^+$ at 4820 \AA\ represents the earliest cometary detection of the cations discussed here \citep{Swings65}.  The spectra of the tails from comet C/1907 L2 (Daniel) and the exceptionally bright comet C/1908 R1 (Morehouse) both yielded correspondence to emission bands that were soon matched to discharge spectra acquired in the laboratory \citep{Fowler1909}. Owing to the low sublimation temperature of CO,  CO$^{+}$ emission can be seen in comet spectra at distances exceeding 5 AU \citep{cochran91}.  CO$^+$ has subsequently been utilized as a spectral reference point in the cometary detection of CO$_2^+$ and OH$^+$ \citep{Swings50}.

Extrasolar rotational lines of CO$^+$ were first observed towards the photodissociation region M17SW in 1993 \citep{Latter93}, further showcasing the near-century lag between comet cation chemistry and interstellar cation chemistry. CO$^+$ has since been observed towards other photodissociation regions, planetary nebulae, in the disk of M82, and in the circumnuclear torus of Cygnus-A \citep{Fuente2000,Fuente2006, Bell2007, Stauber2009}.  Even though CO$^+$ had been known in comets previously from near-UV to visible spectra, the established $J=2 \rightarrow J=1$ rotational line of CO$^+$ was also observed in comet Hale-Bopp with the Caltech Submillimeter Observatory in 1997 \citep{lis97a, lis97b}.  However, more modern spectral benchmark data would greatly enhance the spectral resolution needed to inform more detailed future observations.

\begin{table}
\begin{center}
\caption{\label{co+} Reported Vibrational Frequencies (cm$^{-1}$) and Rotational Constants (MHz) for the Various Electronic States of CO$^+$.$^a$}   
\begin{tabular}{llcc}
State  & & $\nu=0$ & $\nu=1$ \\
\hline                    
$X\ ^2\Sigma^+$ & $h\nu$ & 0.0 & 2183.9 \\
 & $B$       & 59 270.5 & \\
 & $D$       & 0.190 &  \\
\hline
$A\ ^2\Pi$ & $h\nu$ & 20733.3 & 1534.9 \\
 & $B$       & 47 649 &    \\
 & $D$       & 0.20 &  \\
 \hline
$B\ ^2\Sigma^+$ & $h\nu$ & 45876.7  & 1678.3 \\
 & $B$       & 53 930 &   \\
 & $D$       & 0.23 &   \\
  
\hline
\end{tabular}

\end{center}
$^a$ Data collected from \cite{Irikura07, Hakalla19}.
\\

\end{table}

There is an extensive body of work reporting on spectral data of CO$^+$ that has spanned multiple decades with most of the modern foundation coming the past 40 years \citep{Marchand69, Brown84, Kuo86, Bembenek94, Haridass92}. The current state of knowledge is listed in Table \ref{co+}.  Some of the most complete examinations for the spectral classification of CO$^+$ are from  observations of cometary ion tails \citep{vujisic88, Haridass00, kepa04} and from quantum chemical explorations \citep{lavendy93, xin18}.  A very recent laboratory-based Fourier-transform emission spectral examination of this simple molecule has revealed an extensive trove of rovibronic data for the $A\ ^{2} \Pi_{i} \rightarrow X\ ^{2} \Sigma^{+}$ transition \citep{Hakalla19}.  The experimental data are fit from previous known values, including additional reference points from the $B\ ^{1}\Sigma^+ \rightarrow A\ ^{1} \Pi$ transition of neutral carbon monoxide.  In turn, these results have been used to provide input parameters for spectroscopic modeling with the PGOPHER software \citep{western10}.  The remaining, highly-resolved ($\sim1$ cm$^{-1}$) experimental lines were then matched with the output from PGOPHER such that the rest of the $A\ ^{2} \Pi_{i} \rightarrow X\ ^{2} \Sigma^{+}$ rovibronic features within 2000 cm$^{-1}$ of the onset are classified. Hence, the data necessary to model CO$^+$ have been produced and are curated within this paper but are not accessible in a standard format beyond that which is listed as data tables.

 \citet{magnani86} performed a comprehensive series of calculations on the fluorescent equilibrium of CO$^{+}$.  Unlike prior studies, they included the Baldet-Johnson bands and the first negative bands. The data used for their calculations were based on a compilation of nearly two dozen empirical and theoretical papers. Line positions (given in cm$^{-1}$) were taken from \citet{schmid1933}, \citet{Bulthuis1934}, \citet{rao1950b}, and \citet{herzberg1950}. Oscillator strengths were drawn from \citet{joshi1966} and \citet{Jain1972}.  The Frank-Condon factors were taken from \citet{nicholls1962}, and H\"{o}nl-London factors were utilized from \citet{rao1950a}, \citet{rao1953}, and \citet{schadee1964}.  Transition probabilities were based on \citet{arpigny1976} and \citet{crovisier1985}.  Information on transitions were taken from \citet{kopelman1962}, \citet{arpigny1964a}, \citet{arpigny1964b}, \citet{certain1973}, \citet{Feldman1976}, \citet{rosmus1982}, and \citet{chin1984}. Lab measurement data from \citet{lawrence1965} and \citet{judge1972} were also used.  Formulas for the Einstein A and B coefficients were taken from \citet{schleicher1982}.  Solar spectral data for the full solar disk were taken from \citet{broadfoot72} and \citet{ahearn1983}.  Because the oscillator strength for the Baldet-Johnson bands was unknown, \citet{magnani86} assigned a value such that transition probabilities were in agreement with available laboratory observations, resulting in projected band intensities that were uncertain but likely still reliable at the 20$\%$ level and consistent with their non-detection in most comets. Tables are provided in their paper with the fluorescence efficiencies of 60 transitions for a range of heliocentric velocities between -50 to +350 km/s. However, as explained in our discussion above, significant work has been done on the spectrum of CO$^{+}$ after the \citet{magnani86} paper was published (see Table 4 for recommended constants and frequencies). These recent studies, such as \citep{xin18, Hakalla19}, have increased the spectral resolution as well as the accuracy of known transitions. As such, the fluorescence model of CO$^+$ would benefit from including these spectral models.

\subsection{OH$^+$}

The  $A\ ^2\Sigma^+ - X\ ^2\Pi_i$ emission of neutral OH around 3085 \AA\ is readily used to derive water production rates of comets \citep[cf.][]{ahearn1995}, and in the ISM, the hydroxyl radical was actually the first molecule to be observed via rotational spectroscopy in 1963 towards the very powerful radio source in the Cas A supernova remnant \citep{Weinreb63} and has since been observed in diffuse clouds, as well \citep{Neufeld10, Porras13}. Both OH$^+$ and CO$_2^+$ were observed first in comet C/1947 F1 (Bester) \citep{Swings50}.   The dominant feature for this OH$^+$ observation is what spectroscopists now label as the $A\ ^3\Pi \rightarrow X\ ^3\Sigma^-$ transition between 3300 -- 3600 \AA. This broad emission feature may contaminate observations acquired with comet narrowband filters designed for continuum measurements or for the emission of NH around 345 nm \citep{Bodewits2016}. Outside comets, OH$^+$ was not observed until 2010, toward Sgr B2 \citep{Wyrowski10} and background stars at near-UV wavelengths \citep{Krelowski10}.  OH$^{+}$ has also been observed around ultraluminous galaxies \citep{vdw2010}, toward a lensed quasar \citep{muller2016}, in the Orion bar \citep{vdt2013}, and in cometary knots of planetary nebulae \citep{aleman14, priestley18}. 

\begin{table}
\begin{center}
\caption{\label{oh+} Reported Vibrational Frequencies (cm$^{-1}$) and Rotational Constants (MHz) for the Various Electronic States of OH$^+$.$^a$}   
\begin{tabular}{llcc}
State  & & $\nu=0$ & $\nu=1$ \\
\hline                    
$X\ ^3\Sigma^-$ & $h\nu$ & 0.0 & 2956.358469 \\
 & $B$       & 492 346.37 &  470 531.9\\
 & $D$       & 57.609 9   &  56.014 5 \\
 & $H$       & 0.003 942  &  0.003 903\\
\hline
$A\ ^3\Pi$ & $h\nu$ & 27935.6930 & 1975.9872 \\
 & $B$       & 400 841.1 & 375 144.7 \\
 & $D$       & 68.056    & 65.885 4  \\
 & $H$       & 0.004 350 &  0.003 007\\
 \hline
  
\hline
\end{tabular}
\end{center}
$^a$ Data collected from \cite{Hodges2017}.
\\
\end{table}

Like with CO$^+$, the foundational experimental spectroscopic work on OH$^+$ comes from the previous 40 years or so \citep{Merer75, Bekooy85, Gruebele86, Liu87, Varberg87, Rehfuss92}.  During this same era, \citet{saxon86} conducted a theoretical study of OH$^{+}$ photodissociation from the ground state and calculated potential curves for each of the lowest three $\Sigma^{-}$ and $\Pi$ states.  They also evaluated transition dipole moments between the ground and excited states and used them to calculate photodissociation cross sections.  \citet{gc2014} conducted a theoretical study of OH$^{+}$ which included calculating the Einstein coefficients for ro-vibrational bands involving the X$^3\Sigma^-$ and $A\ ^3\Pi$ electronic states and calculating the state-to-state rate constants for inelastic collisions between He and OH$^{+}$ (X$^3\Sigma^-$).  A more recent study by \cite{Hodges2018} combines empirical energy surfaces with the Rydberg-Klein-Rees method, dipole moment calculations calculated quantum chemically, and oscillator strengths and Einstein A coefficients calculated using PGOPHER. Line lists are created with positions, oscillator strengths, and Einstein A coefficients for the $A\ ^3\Pi \rightarrow X\ ^3\Sigma^-$ rovibronic transition and the rovibrational transitions of the X$^3\Sigma^-$ ground state, including the following levels: (A, $\nu=0$, 1; X, $\nu=0, 1, 2, 3, 4$) up to $J''=30$. The line positions of the $A\ ^3\Pi \rightarrow X\ ^3\Sigma^-$ band (3300 -- 3600 \AA) are compared to those determined using a discharge cell combined with a Fourier transform spectrometer \citep{Hodges2017}. The authors state that there is no intensity measurement to verify the vibrational transitions (around 3.4 $\mu m$) of the {\em X} -- {\em X} band. A table of their data is available electronically through the VizieR database, and the pertinent rovibronic features of OH$^+$, are listed in Table \ref{oh+}.

\subsection{O$_2^+$}

Surprisingly, neutral molecular oxygen was detected at relatively large abundances with the ROSINA and Alice instruments on board \emph{Rosetta} as it orbited comet 67P/Churyumov–Gerasimenko \citep{Bieler15, keeney17}. It is unclear whether this O$_2$ was present at the comet's formation,  is trapped in cometary ices or clathrates \citep{LuspayKuti2018}, or is produced by a chemical reaction with materials on the surface or in the coma \citep{Fortenberry2019}.  Albeit its presence in cometary atmospheres was first suggested 70 years ago \citep{Swings50}, the molecular oxygen cation actually has not been detected remotely to date in either cometary or interstellar media \citep{Glinski2004}, and was only spuriously detected in situ by \emph{Rosetta}'s ROSINA instrument \citep{Beth2020}.  Three faint rotational lines of $X\ ^3\Sigma_g^-$ O$_2$ were finally detected in 2011 towards Orion \citep{Goldsmith11}, but the cation remains elusive for detection in the ISM as well as in comets. In comets, photoionization rates for the production of O$_2^+$ from O$_2$ are  larger ($\sim 10\%$) than for the production of H$_2$O$^+$ by photoioniziation of H$_2$O ($\sim3 \%$; \citet{Huebner2015}), implying that  O$_2^+$ should be present as well, especially outside the collisionaly thick inner coma where it can quickly dissipate through chemical reactions \citep{Beth2020}.  However, high-quality spectral reference data and fluorescence efficiency rates are needed to look for its signature in cometary and other astrophysical spectra. 

\begin{table}
\begin{center}
\caption{\label{o2+} Reported Vibrational Frequencies (cm$^{-1}$) and Rotational Constants (MHz) for the Various Electronic States of O$_2^+$.$^a$}   
\begin{tabular}{llcc}
State  & & $\nu=0$ & $\nu=1$ \\
\hline                    
$X\ ^2\Pi_g$ & $h\nu$ & 0.0 & 1872.27 \\
 & $B$       & 50 704 & \\
 & $D$       & 0.159  & \\
\hline
$A\ ^2\Pi_u$ & $h\nu$ & 40669.3 & 871.11 \\
 & $B$       & 31 829  &  \\
 & $D$       & 0.178  &  \\
 \hline
  
\hline
\end{tabular}
\end{center}
$^a$ Data collected from \cite{Irikura07}.
\\
\end{table}

\citet{li2000} present results of laser-induced fluorescence (LIF) of the (8,0) and (8,1) bands of the $A\ ^2\Pi_u \rightarrow X\ ^2\Pi_g$ system (between 1800 -- 5300 \AA), as well as preliminary LIF determinations of ($\nu$ = 0, 1) distributions from reactions involving Ar$^{+}$ and Xe$^{+}$.  Electron impact ionization-induced spectra between 2200 -- 6000 \AA\ are presented in \cite{Terrell2004}, and these are gathered in Table \ref{o2+}. They present a simplified molecular model to interpret their data, which includes $A\ ^2\Pi_u \rightarrow X\ ^2\Pi_g$ and $b\ ^4\Sigma^-_g \rightarrow a\ ^4\Pi_u$ (between 4500 -- 8500 \AA).  All other excitations are outside the window of solar radiation observed from the ground \citep{NIST2}, and spin-flip excitations are not likely to produce any measurable intensity under cometary conditions.  The models of these band features were made possible by the rotational constants and strong classification of the vibrational fundamental from \cite{NIST2}.  The fundamental frequency varies significantly across these four states from 898.2 cm$^{-1}$ in $A\ ^2\Pi_u$ \citep{Colburn77} to 1904.7 cm$^{-1}$ in the ground $X\ ^2\Pi_g$ state, where these bookend the 1580.19 cm$^{-1}$ fundamental in the $X\ ^3\Sigma^-_g$ in the neutral \citep{NIST2}.  The rotational constants vary from 31829 MHz to 50704 MHz, again respective of $A\ ^2\Pi_u$ and $X\ ^2\Pi_g$  \citep{Irikura07}.  Consequently, the rovibronic lines have notably different progressions within them in the doublet excitation of the shorter wavelengths.

\section{Conclusions}

The state of available spectral data for small, oxygen-containing molecular cations is mixed but is rife with notable knowledge gaps that hampers the full diagnostic use of the emission of molecular ions.  These holes must be filled in order to better understand physical processes and chemical evolution for tenuous atmospheres  of various astronomical bodies, such as comets around perihelion.  General near-UV to visible spectral progressions are largely established for these molecules, and the rovibrational nature of the ground electronic states have been well-classified for these relatively simple molecules.  However, a rigorous, quantitatively-predictive analysis of the full rovibronic structure, especially the rotational constants of the excited rovibronic states, is currently lacking in most cases. Existing excitation models are decades old, and rely on limited spectral models. This work shows where the data in these models can and should be improved.

The worst case of the current state of knowledge of the spectral properties of the H$_2$O$^+$, CO$_2^+$, CO$^+$, OH$^+$, and O$_2^+$ set is the water cation, where only low-resolution characterization for the rovibrational transitions of the two lowest electronic states is available in the literature currently.  These available data are also mostly limited to the vibrational band positions and have no rotational substructure described.  While surprising for a molecular product of such a common and abundant molecule, this represents a readily-surmountable challenge for spectroscopic characterization for the astrochemical community for both high-resolution experiment and modern quantum chemistry, if not in a combination of the two approaches.  The reference data for CO$_2^+$ and CO$^+$ have similar rotational spectral holes, but the vibronic spectra of the higher electronic states have been explored at high resolution for the most part.  In any case, these small, oxygen-containing molecular cations and their daughter species are often observed in bodies such as comets, but clear identification of the lines, especially the rovibronic features, is necessary in order to gain deeper insights into the photochemistry playing out in various Solar System environments.

\section{Acknowledgements}
RCF acknowledges funding from NASA Grant NNX17AH15G, NSF Grant OIA-1757220, and start-up funds provided by the University of Mississippi.  DB acknowledges support from NASA Grant 80NSSC19K1304 for the `Rosetta Data Analysis Program'.  DMP acknowledges support from the Southeastern Conference Faculty Travel Program.

\bibliography{refs}
\bibliographystyle{aasjournal}
\end{document}